\documentclass{article}
\usepackage{graphicx}  
\usepackage{amsmath}   
\usepackage{amssymb}   
\usepackage{bm} 
\usepackage{dcolumn}
\usepackage{color}
\usepackage{mathrsfs}
\usepackage{amsfonts}
\usepackage{varioref}
\RequirePackage[colorlinks,citecolor=blue,urlcolor=magenta,linkcolor=blue]{hyperref}
\labelformat{section}{Section #1} 
\labelformat{subsection}{Section #1} 
\labelformat{subsubsection}{Section #1}
\labelformat{subsubsubsection}{Section #1}
\labelformat{equation}{Eq.~(#1)} 
\labelformat{figure}{Fig.~#1} 
\labelformat{subfigure}{Fig.~\thefigure#1} 
\labelformat{table}{Tab.~#1} 
\labelformat{appendix}{Appendix #1}

\title{\bf Cosmic structure sizes in generic dark energy models}
\author{\bf  Sourav Bhattacharya$^{2}$\footnote{sbhatta@iitrpr.ac.in} and Theodore N Tomaras$^{2}$\footnote{tomaras@physics.uoc.gr}\\
$^{1}$\small{Department of Physics, Indian Institute of Technology Ropar, Rupnagar, Punjab 140 001, India}\\
$^{2}$\small {ITCP and Department of Physics, University of Crete, 700 13 Heraklion, Greece } }

\begin{document}
  
\maketitle

\begin{abstract}
{\noindent  The maximum allowable size of a spherical cosmic structure as a function of its mass is determined by the maximum turn around radius $R_{\rm TA,max}$, the distance from its centre where the attraction on a radial test particle due to the spherical mass is balanced with the repulsion due to the ambient dark energy. In this work, we  extend the existing results in several directions. (a) We first show that for $w\neq -1$, the expression for $R_{\rm TA, max}$ found earlier using the cosmological perturbation theory, can be derived using a static geometry as well.   (b) In the generic dark energy model with arbitrary time dependent state parameter $w(t)$, taking into account the effect of inhomogeneities upon the dark energy as well, where it is shown that the data constrain $w(t={\rm today})>-2.3$, and (c) in the quintessence and the generalized Chaplygin gas models, both of which are shown to predict structure sizes consistent with observations.     
}
\end{abstract}
  

\maketitle
\section{Introduction and overview}
\noindent
Since the discovery of the accelerated expansion of our current Universe at the end of last century, the dark energy, along with the cold dark matter have become the central themes in the theory and applications of modern cosmology. The simplest of the dark energy models is certainly a positive cosmological constant, $\Lambda$ and the simplest model of the modern universe is $\Lambda$-Cold Dark Matter ($\Lambda{\rm CDM}$). The $\Lambda{\rm CDM}$ has so far passed with flying colors all cosmological observation tests ~\cite{Weinberg:2008zzc, Lapuente}, starting from the redshift of type Ia Supernovae, the Hubble rate, the galaxy clustering, the microwave background radiation and so on \footnote{For recent criticism, however, of the statistical significance of the Supernovae Ia data see ~\cite{sarkar} and references therein.}.  

Since the observed value of $\Lambda\sim {\mathcal O}(10^{-52}{\rm m}^{-2})$ is tiny compared to the inverse of the Planck length squared, it is natural to expect that its effect would be observable at large space-time scales only, as for instance imprinted in the light coming from some high redshift supernovae or in the data for the early universe. However, a novel potentially local check of the dark energy was proposed recently~\cite{PavlidouTomaras2013, PavlidouTetradisTomaras2014, TanoglidisPavlidouTomaras2014}, pertaining the stability of large scale structures. The idea is simple, based on the observation that the maximum size of an e.g. spherical bound structure should be the distance from its center at which the attractive Newtonian force on a test mass due to the spherical overdensity is balanced with the repulsion due to the ambient dark energy. Beyond this distance, which was called ``maximum turnaround radius" and labeled $R_{\rm TA,max}$ a test particle cannot stay bound, but will be dragged away by the antigravity effect of the dark energy. The same idea was used in \cite{weinberg2} to derive an upper bound on a hypothetical at that time  cosmological constant on the basis of the existence of galaxies. The next step is then to use $R_{\rm TA,max}$ as an observable to constrain the parameters of any cosmological model of interest, by comparing its theoretical prediction for $R_{\rm TA,max}$ with the actual data. 

\begin{figure}[ht]
\begin{center}
\includegraphics[width=.8\textwidth, clip=true]{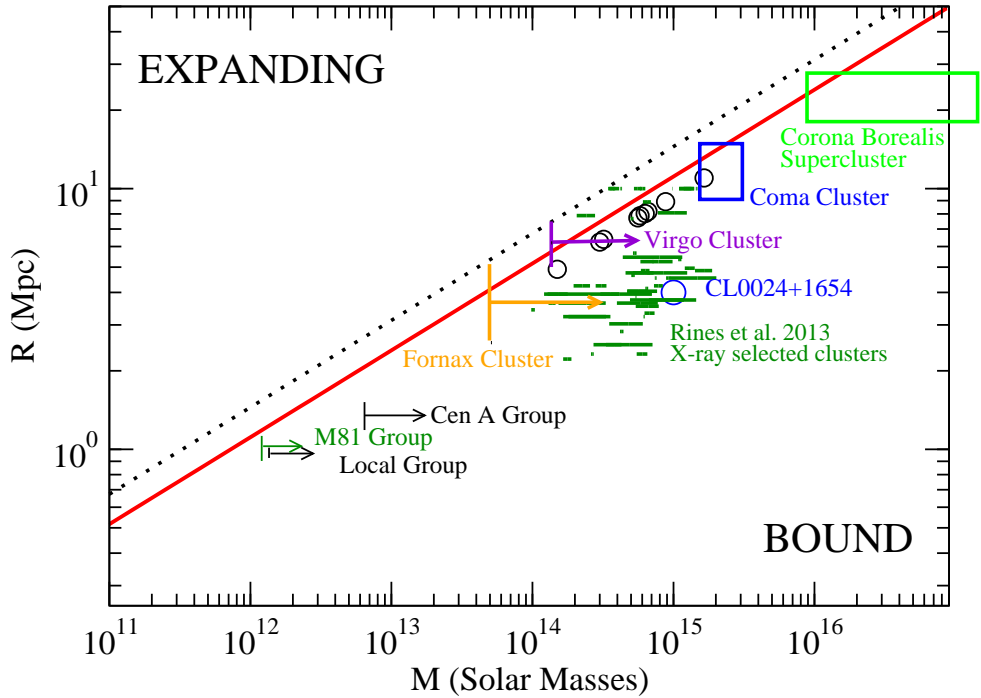}
\end{center}
\caption{\label{fig} Observed size versus mass plot for several nearby cosmic structures compared to the $R_{\rm TA,max}$ predicted by $\Lambda{\rm CDM}$ (red line). The dotted line represents the maximum departure due to non-sphericities. ({\small Taken from Ref.~\cite{PavlidouTomaras2013}, $\copyright$ SISSA Medialab Srl.  Reproduced by permission of IOP Publishing.  All rights reserved.)}}
\end{figure}

For $\Lambda{\rm CDM}$, in particular, the predicted value of $R_{\rm TA,max}$ for a spherical structure equals $(3M G/\Lambda c^2)^{1/3}$, the details of which would be presented in the next sections. \ref{fig}~\cite{PavlidouTomaras2013} (see also references therein), shows the size versus mass of some nearby large scale structures, together with the theoretical prediction of the model. As is evident, for superclusters as large as $M\gtrsim 10^{15}M_{\odot}$, the prediction of $\Lambda{\rm CDM}$ lies very close from above, with the departure from the data being roughly only about $10\%$! Thus, the $\Lambda{\rm CDM}$ model is absolutely consistent with the stability of cosmic structures from the maximum turn around perspective \footnote{Note in~\ref{fig} that the Corona Borealis supercluster seems to have considerable effect due to non-sphericity. However, it may not be an integral structure at all, but a giant binary connected by a filament  like structure, see ~\cite{PavlidouTomaras2013, Pillastrini:2015oqa} and references therein.}. It is worth noting here that the structures we are looking into are very close to us ($z\ll 1$), and hence this is a local check or challenge to the dark energy -- the dark energy is at work right within the structure itself.
%
 After a certain structure decouples from the expansion of the surrounding Universe, it shrinks and virializes by redistributing the kinetic energies and positions of its individual constituents~\cite{Padmanabhan1993} under the action of the attractive force due to its mass, which is now dominant. Given that the process of virialization of a structure typically enhances any non-sphericity in its initial profile, a spherical model for a structure is more appropriate when it is away from virialization. Furthermore, it was
shown using the Press-Schechter mass function in~\cite{TanoglidisPavlidouTomaras2014, TanoglidisPavlidouTomaras2016} that there exists a transitional mass scale $\sim 10^{13}M_{\odot}$, above which the structures are {\it not} virialized today.  Thus, in order to challenge a dark energy model using the maximum size versus mass criterion, it is most effective to look into structures which lie above that transitional mass scale. We refer our reader to~\cite{Einasto:2016rhe} for a discussion on the future evolution of the cosmic webs of such large masses and to~\cite{Lee:2015upn, Lee2016, Lee:2016oyu} for a discussion on the possible violation of the maximum turn around bound by looking into the peculiar velocity profiles of members of a given structure. We further refer our reader to e.g.~\cite{Stuclik1, Stuclik2} on aspects of particle motion in the black hole Schwarzschild-de Sitter spacetime relevant to the derivation of $R_{\rm TA,max}$ in $\Lambda{\rm CDM}$ and to~\cite{Stuclik3, Stuclik4, Stuclik5} for effects of a positive $\Lambda$ on astrophysical scenarios or phenomena such as the accretion disks. 

Despite the simplicity and the most overwhelming successes of the $\Lambda{\rm CDM}$ model, it is well known that it suffers from the so called ``fine tuning of the cosmological constant" problem, the fact that $\Lambda$, when interpreted and computed as the vacuum energy density, most outrageously disagrees with its actual observed value today~\cite{Martin2012}. Furthermore, it does not provide any insight towards an explanation of the ``coincidence problem", the fact that the current numerical values of the vacuum and matter energy densities are so close to each other. Attempts to resolve these issues on the basis of the instability of de Sitter space~\cite{Tomaras1986, Floratos1987, Polyakov2012, Woodard2014} are not yet entirely satisfactory. Finally, in~\cite{Sahni2014}, a possible discrepancy between the $\Lambda{\rm CDM}$ model and the Baryon Acoustic Oscillation data for high redshift has been proposed. All these issues and also the so far lack of any observational evidence of a dark matter particle candidate, have triggered in recent times vigorous research for alternatives of the $\Lambda{\rm CDM}$ model and/or Einstein's theory of gravitation~\cite{CliftonEtAl2012, BullEtAl2015, AlonsoEtAl2016} (also references therein). As a result, a plethora of potential candidate theories or models have been proposed and the need to invent effective observables to distinguish them has become apparent. See also~\cite{Novikov:2016hrc, Novikov:2016fzd} for a recent proposal of quantum modification of the  General Relativity, by treating the massive gravitons as the primary dark matter.

Since its proposal, the maximum turn around radius as a useful cosmological observable has received considerable attention in the context of modified or alternative gravity/dark energy models~\cite{Faraoni2015}-\cite{Lee:2016bec}. The goal is of course to constrain the additional parameters of such theories. For example, for a McVittie spacetime~\cite{Gao:2004wn, Kaloper:2010ec, AntoniouPerivolaropoulos2016}, derivation of the turn around radius can be found in~\cite{Faraoni2015}-\cite{Faraoni:2015cnd}, by analyzing geodesics and also using the quasilocal mass function of~\cite{Hawking, Hayward}.  In~\cite{BhattacharyaDialektopoulosTomaras2015} a constraint on the matter-galileon coupling parameter was obtained for a cubic galileon model~\cite{Martin-Moruno:2015kaa}. Both an upper and a lower bound was obtained for this coupling. The lower bound is new, while the upper is improved by about $50\%$ from what was known on the basis of the Solar system~\cite{Iorio:2012pv} data. The derivation of $R_{\rm TA,max}$ in the Brans-Dicke theory~\cite{BransDicke1961} in the presence of a positive cosmological constant $\Lambda$ was carried out in~\cite{BhattacharyaEtAl2015, Bhattacharya:2016vur}. In this theory the $R_{\rm TA,max}$ is greater than that of $\Lambda{\rm CDM}$. Qualitatively, this is due to the fact that the extra scalar of the Brans-Dicke model contributes to make the attractive force stronger, thus allowing for larger structures. Consequently, no constraint on its parameter $\omega$ was obtained. Two more models with the same effective behavior will be presented in Section~3.2. 

There are two ways to derive the maximum turn around radius -- one is to analyze geodesics in a static 
geometry and the other is via the study of a test fluid moving in a background cosmological McVittie spacetime~\cite{PavlidouTomaras2013, PavlidouTetradisTomaras2014}.  It has been explicitly demonstrated  that both these approaches yield the same result for the $\Lambda{\rm CDM}$ model~\cite{PavlidouTetradisTomaras2014}, as well as for the Brans-Dicke theory with a positive $\Lambda$~\cite{Bhattacharya:2016vur}. \\

\noindent
Let us now emphasize the new results we derive in the rest of this  paper. For $w\neq -1$, the dark energy is taken to be time dependent and as we mentioned earlier, for a constant such $w$, the expression for $R_{\rm TA, max}$ was derived earlier in~\cite{PavlidouTetradisTomaras2014} using cosmological scalar perturbation theory. Can we have a derivation of the same result using a static geometry as well, just like $w=-1$? The answer turns out to be, Yes.
In Sec.~2, we consider in the context of General Relativity a static and spherically symmetric dark energy fluid with equation of state $P(r)=w\rho(r)$ with constant $w$. We derive a first order metric using this source and  the maximum turn around radius using this metric is found to match exactly with the one derived in~\cite{PavlidouTetradisTomaras2014}.

In many alternative  dark energy models, the state parameter, $w$, could be time dependent. Therefore, it is important to have a derivation of $R_{\rm TA,max}$ in the context of such dark energy models as well. In Sec~3.1 we shall present a derivation of the $R_{\rm TA,max}$ in the cosmological scenario for a dark energy with an {\it arbitrary, time dependent} state parameter $w(t)$  ($P_E(t)=w(t)\rho_E(t)$) and by also taking into account the effect of inhomogeneities upon the dark energy distribution. We shall further apply in Subsection 3.2 this generic result in the context of the quintessence, as well as of the generalized Chaplygin gas models, establishing their consistency with the observational data. We conclude in Section 4 with a brief discussion and outlook.

We shall use mostly positive signature $(-,+,+,+)$ for the metric and will set $c=1$ throughout.

\section{$R_{\rm TA,max}$ in the static geometry }
\noindent
It is well known that the positive $\Lambda$ permits a static and spherically symmetric solution within the cosmological event horizon, namely, the Schwarzschild-de Sitter spacetime,
\begin{eqnarray}
ds^2=-(1-2MG/r-H_0^2r^2)dt^2+ (1-2MG/r-H_0^2r^2)^{-1}dr^2+r^2 d\Omega^2
\label{ta0}
\end{eqnarray}
where $d\Omega^2=d\theta^2+\sin^2\theta d\phi^2$ is the unit sphere line element and $H_0=\sqrt{\Lambda/3}$. Let us consider a timelike geodesic $u^a$ ($u_au^a=-1$) in this spacetime. The above spacetime is endowed with one timelike and three spacelike Killing vector fields generating a 2-sphere,
\begin{eqnarray}
\zeta_0^a&&=(\partial_t)^a \qquad \zeta_1^a= -\sin\phi (\partial_{\theta})^a-\cot\theta \cos\phi (\partial_{\phi})^a\nonumber\\
\zeta_2^a&&= \cos\phi (\partial_{\theta})^a-\cot\theta \sin\phi (\partial_{\phi})^a \qquad \zeta_3^a=(\partial_{\phi})^a
\label{ge1}
\end{eqnarray}
The conserved quantities along the geodesic corresponding to each of these Killing vector fields are given by,
\begin{eqnarray}
E&&=-g_{ab}u^a\zeta_0^b= (1-2MG/r-H_0^2r^2)\dot{t} \qquad L_1= g_{ab}u^a\zeta_1^b= -r^2\left(\dot{\theta}\,\sin\phi+\dot{\phi}\, \sin\theta \cos\theta \cos\phi\right)\nonumber\\
L_2&&=g_{ab}u^a\zeta_2^b=r^2\left(\dot{\theta}\,\cos\phi-\dot{\phi}\, \sin\theta \cos\theta \sin\phi\right) \qquad   L_3 = g_{ab}u^a\zeta_3^b= \dot{\phi}\,r^2\sin^2\theta 
\label{ge2}
\end{eqnarray}
where the `dot' denotes differentiation with respect to the proper time along the trajectory. The above conserved quantities are respectively identified as the conserved energy and the conserved components of the orbital angular momentum of the geodesic. Expanding the on-shell condition $u_au^a=-1$ in the background of~(\ref{ta0}) and using Eqs.~(\ref{ge2}), we have
\begin{eqnarray}
\dot{r}=\pm \sqrt{E^2-\left(1-2MG/r-H_0^2r^2\right)\left(L^2/r^2+1\right)}
\label{ge3}
\end{eqnarray}
where $L^2=L_1^2+L_2^2+L_3^2=r^4(\dot{\theta}^2+\dot{\phi}^2\,\sin^2\theta )$ and the $\pm$ sign denotes outgoing and incoming trajectories, respectively. The maximum turn around is determined by the condition of vanishing acceleration $\ddot{r}=0$, or,
\begin{eqnarray}
\pm\frac{L^2}{r^3}\left(1-2MG/r-H_0^2r^2\right)\mp\left(L^2/r^2+1\right)\left(MG/r^2-H_0^2r\right)=0
\label{ge4}
\end{eqnarray}
It is straightforward to convince oneself that, as expected, the consequence of angular momentum is to decrease the size of a stable structure with a given mass, because the attraction of the central mass has to counterbalance the additional centrifugal force due to the rotation. Indeed, define
$$
R_0^3=\frac{M}{H_0^2},\; \lambda^2=\frac{L^2}{MR_0}, \; m=\frac{3M}{R_0}=3(MH_0)^{2/3} \;\; {\rm and} \;\; x=\frac{r}{R_0}
$$ 
and write the condition (\ref{ge4}) in the form
\begin{equation}
\lambda^2\left(1-\frac{m}{x}\right) = x(1-x^3)
\label{ge5}
\end{equation}
The cosmic structures of interest have sizes much larger than their Schwarzschild radius and much smaller than the cosmological horizon, i.e. $2M\ll r \ll H_0^{-1}$ or equivalently
$
m\ll x \ll m^{-1/2},
$
with $m\ll 1$ for consistency. For $\lambda=0$, (\ref{ge5}) has the well known solution $x=1$ ($r=R_0=(M/H_0^2)^{1/3}$). Introducing a small angular momentum $\lambda^2={\mathcal O}(\epsilon)$ the above solution changes to 
$
x=1-\frac{1}{3} (1-m) \lambda^2 < 1,
$
which confirms that the angular momentum diminishes the size of the cosmic structures. The above argument can be generalized numerically to all physical angular momenta.

Thus we conclude that  the maximum possible size of a structure with a given mass is the maximum turn around radius for $L=0$, that is     
\begin{eqnarray}
R_{\rm TA,max}= \left(\frac{3MG}{\Lambda}\right)^{\frac13}
\label{ge5}
\end{eqnarray}
In other words, this is precisely the point where the attraction due to the central mass gets balanced with the repulsion due to the dark energy, for radial timelike geodesics. 

The above derivation could also be performed using the orbits of the timelike Killing vector field of~(\ref{ta0}). Precisely, the static observers along these timelike Killing vector field would feel a spacelike acceleration $a^b=(\partial_t)^a\nabla_a(\partial_t)^b=\left(GM/r^2-H_0^2r \right)\nabla^b r$. Thus the acceleration vanishes at $r=R_{\rm TA,max}$, corresponding to the maximum of the norm of the timelike Killing vector field, giving the maximum size of the structure. We also note here that 
for structures up to as large as $M\sim 10^{16} M_{\odot}$ and taking $\Lambda\sim 10^{-52}{\rm m}^{-2}$, the $(2MG/r+H_0^2 r^2)$ term appearing in the metric functions is much less than unity at this length scale. In other words, the maximum we are dealing with here is a very broad maximum for practical astrophysical scenarios. 

Even though the above computation might seem idealistic as the framework used was static, it can be generalized, as we shall see in the next section, to include the $\Lambda{\rm CDM}$ cosmological perturbation theory and leads to exactly the same answer~\cite{PavlidouTetradisTomaras2014}, as expected from the possible gauge or coordinate invariance of the maximum turn around radius~\cite{Velez:2016shi}. We shall also further extend this result there with general dark energy with equation of state $P_E(t)=w(t)\rho_E(t)$.

Staying at the constant $w$ case, for $w\neq -1$, the dark energy is in general treated as time dependent and spatially homogeneous. For a constant such $w$, the maximum turn around radius was found to be $\left(-3M G/(4\pi(1+3 w)\rho_E)\right)^{1/3}$~\cite{PavlidouTetradisTomaras2014}. For $w=-1$, we may identify $\Lambda=8\pi G \rho_E$ to recover the $\Lambda{\rm CDM }$ result. In the following, we wish to present a novel alternative derivation of this result in the framework of a static geometry, assuming accordingly that the dark energy is static as well. 

The concept of static dark energy was earlier introduced 
in e.g.~\cite{Chapline}-\cite{Bhar:2016dhd} to construct compact and static star-like solutions, known as the ``dark energy stars", made up of a dark energy fluid. This corresponds to the assumption that in the vicinity of a structure or central mass, the dark energy becomes inhomogeneous too. Quite surprisingly, as we shall see, the answers obtained in these two ways match! However, it is important to emphasize that we are not considering any dark energy stars here, but only generic large scale structures in the ambiance of a static dark energy fluid. Certainly, as we move away from the object, the assumption of staticity would break down.

Let us then begin with the ansatz for a static spherically symmetric spacetime,
\begin{eqnarray}
ds^2=-f(r)dt^2+h(r)dr^2+r^2 d\Omega^2.
\label{ta1}
\end{eqnarray}
We assume that the energy momentum tensor for the dark energy corresponds to that of an ideal fluid,
\begin{eqnarray}
T^E_{ab}=\rho_{E}\, u_au_b+P_E\,\left(g_{ab}+u_a u_b\right),
\label{ta2}
\end{eqnarray}
where $\rho_E$, $P_E$ and $u_a$ are respectively the energy density, pressure and the velocity of the fluid's world line. For a backreacting $T^E_{ab}$ in a static geometry, $u_a$ should be parallel to the orbits of the timelike Killing vector field. If we normalize it to
unity, $u_au^a=-1$,  we may take $u^a=f^{-1/2}(\partial_t)^a$. We assume that the parameter $w$ in the fluid equation of state $P_E= w\rho_E$, is a constant. We shall take $\rho_E>0$, so that $w<-1/3$, necessary to violate the strong energy condition and generate repulsive effects~\cite{Wald:1984rg}.

The three independent Einstein equations for this system are
\begin{eqnarray}
&&\frac{h'}{h^2 r}+\frac{1}{r^2}-\frac{1}{hr^2}=8\pi G \rho_E, \nonumber\\
&&\frac{f'}{hfr}-\frac{1}{r^2}+\frac{1}{hr^2}=8\pi G w \rho_E,\nonumber\\
&&\frac{f''}{2fh}-\frac{h'}{2h^2 r}+\frac{f'}{2fhr}-\frac{f'h'}{4h^2f}-\frac{f'^2}{4f^2 h}=8\pi G w \rho_E,
\label{ta3}
\end{eqnarray}
where a `prime' denotes derivative with respect to the radial coordinate once and where use was made of the equation of state. Using the conservation equation $\nabla^a T^{E}_{ab}=0$, we obtain 
\begin{eqnarray}
\rho_E(r)=\frac{\rho_{E,0}}{f^{\frac{1+w}{2w}}(r)}
\label{ta3'}
\end{eqnarray}
where $\rho_{E,0}$ is an integration constant. The above equation can be thought of as an analogue 
of the time dependence of the energy density in a homogeneous cosmological spacetime, $\rho_E(t)\sim a(t)^{-3(1+w)}$, where the scale factor $a(t)$ is equivalent to the redshift or the Tolman factor 
$f(r)$~\cite{Wald:1984rg}, for a static spacetime.   We could not find an exact solution of~(\ref{ta3})  
corresponding to the above expression for $\rho_E(r)$. However, since we are chiefly interested in the maximum turn around region, we still could find an approximate and linearized solution there as follows. Since we have seen earlier that the maximum turn around radius in a static spacetime corresponds to the maximum of the norm of the timelike Killing vector field ($f'=0$), we may ignore the variation of $\rho_E$ around this point and write $\rho_E \approx \rho_E(r \sim R_{\rm TA, max}) \approx \rho_{E,0}/f(R_{\rm TA, max})\equiv \rho_0 = \Lambda/8 \pi G~({\rm say})$. Then, the solution of (\ref{ta3}) for the metric is
\begin{eqnarray}
ds^2=&&-\left(1-\frac{2M G}{r}-\frac{\Lambda r^2}{3} \right)^{-\frac{1+3w}{2}}\left(\frac{r}{r_0}\right)^{-\frac{3(1+w)}{2}}
\left\vert \frac{r}{r_{H}} -1 \right\vert^{     \frac{9\left(1+w\right)\left(1-\Lambda r_{ H}^2 /3\right)}{2\Lambda \left(r_{ C}-r_{ H}\right)  \left(r_{ H}-r_{ U}\right) }}\times\nonumber\\    
&&
\times \left\vert \frac{r}{r_{C}} -1 \right\vert^{     -\frac{9\left(1+w\right)\left(1-\Lambda r_{ C}^2 /3\right)}{2\Lambda \left(r_{C}-r_{ H}\right)  \left(r_{ C}-r_{ U}\right) }}
\left\vert \frac{r}{r_{ U}} -1 \right\vert^{     -\frac{9\left(1+w\right)\left(1-\Lambda r_{U}^2 /3\right)}{2\Lambda \left(r_{C}-r_{U}\right)  \left(r_{H}-r_{ U}\right) }} dt^2 \nonumber \\
&&+ \left(1-\frac{2M G}{r}-\frac{\Lambda r^2}{3}\right)^{-1}dr^2 +r^2d\Omega^2, 
\label{ta4}
\end{eqnarray}
The parameter $M$ can be regarded as the mass of a central compact object and $r_0$ is an arbitrary length parameter. 
The three other constants $r_{H},~r_{C} $ and $r_{ U}$ are the nonvanishing roots of the equation $1-2M G/r-\Lambda r^2/3=0$. For the physically interesting case $3MG\sqrt{\Lambda}\leq 1$, all of them are real, namely 
\begin{eqnarray}
r_{H}=  \frac{2}{\sqrt{\Lambda}}\cos\left[\frac13 \cos^{-1}\left(3M G\sqrt{\Lambda}\right)+\frac{\pi}{3} \right],~~ r_{C}=  \frac{2}{\sqrt{\Lambda}}\cos\left[\frac13 \cos^{-1}\left(3M G\sqrt{\Lambda}\right)-\frac{\pi}{3} \right].\nonumber\\ 
\label{ta5}
\end{eqnarray}
are precisely the horizon lengths of the Schwarzschild-de Sitter spacetime, while the third $r_U=-(r_H+r_C)$ is clearly unphysical. However, for $w\neq -1$ they are not known to have  any such physical meaning.
 
Note that for $w=-1$ in \ref{ta4}, we recover, as expected, the Schwarzschild-de Sitter spacetime. Also, in the limit $M\to 0$, it is easy to see that $r_{H} \to 0$ and $r_{C}\to \sqrt{\frac{3}{\Lambda}}$, in which case \ref{ta4} becomes
\begin{eqnarray}
ds^2=&&-\left(1-\frac{\Lambda r^2}{3} \right)^{-\frac{1+3w}{2}} dt^2+ \left(1-\frac{\Lambda r^2}{3}\right)^{-1}dr^2 +r^2 d\Omega^2, 
\label{ta6}
\end{eqnarray}
Finally, if we set $\Lambda=0$ in \ref{ta4}, we recover the Schwarzschild spacetime.

However, since the metrics \ref{ta4} or \ref{ta6} were derived using the  {\it a priori} assumption that we are in the maximum turn around region, to be consistent, we have to linearize it now as follows. 
First we note that since the observed value of the dark energy is tiny, it is reasonable to assume that $MG\sqrt{\Lambda}\ll1$. Then it turns out that $r_{H}\approx 2MG$ and $r_{C}\approx \sqrt{ \frac{3}{\Lambda}}$ in \ref{ta5}. Moreover, the maximum turn around length scale  must be much smaller than the Hubble horizon scale $\sim {\cal O} (\Lambda^{-1/2})$ and also much larger than the Schwarzschild radius of the central mass, i.e. both terms $GM/r$ and $\Lambda r^2/3$ must be much smaller than unity at or around this length scale.  Putting these all in together, we find, up to linear order the metric in \ref{ta4} can be written around the maximum turn around region as,
\begin{equation}
ds^2\approx-\left(1-\frac{2M G}{r}+ \frac{(1+3w)\Lambda r^2}{6} \right)\left(\frac{r_0}{2MG}\right)^{\frac{3(1+w)}{2}}dt^2+ \left(1+\frac{2M G}{r}+\frac{\Lambda r^2}{3}\right)dr^2 +r^2 d\Omega^2
\label{ta7}
\end{equation}
Rescaling the time coordinate $t' =(r_0/2MG)^{3(1+w)/4}t $, and dropping off the prime without any loss of generality, we obtain finally
\begin{eqnarray}
ds^2\approx-\left(1-\frac{2M G}{r}+ \frac{(1+3w)\Lambda r^2}{6} \right)dt^2+ \left(1+\frac{2M G}{r}+\frac{\Lambda r^2}{3}\right)dr^2 +r^2d\Omega^2,
\label{ta7'}
\end{eqnarray}
which for $w=-1$ leads to the linearized Schwarzschild-de Sitter spacetime with $\Lambda$ identified with the cosmological constant. 

As we discussed earlier, the maximum turn around radius in a static spacetime is equivalently given by the maximum of the norm of the timelike Killing vector field
(i.e., $\partial_r g_{tt}=0$), giving
\begin{eqnarray}
R_{\rm TA,max}=\left(-\frac{6MG}{(1+3w)\Lambda}\right)^{\frac13}=\left(-\frac{3M G}{4\pi(1+3 w)\rho_0}\right)^{\frac13},
\label{ta8}
\end{eqnarray}
where in the last equality we have substituted back $\Lambda=8\pi G \rho_{E,0}/f(R_{\rm TA, max})\equiv 8\pi G\rho_0$. This exactly matches 
with the result of~\cite{PavlidouTetradisTomaras2014}, obtained by using the cosmological perturbation theory, and rules out {\it all} dark energy models with $w\lesssim -2.3$. For the cosmological case, $\rho_0$ was to be understood as the dark energy density today for nearby structures. Finally, as a consistency check of \ref{ta7}, it is easy to see that in the maximum turn around region for structures as large as up to
$M\sim 10^{16} M_{\odot}$, both $MG/r$ and $\Lambda r^2$ terms are much smaller than unity, justifying further the simplification $\rho_E(r)\approx \rho_{0}$ we used for the derivation.\footnote{As an aside, we note here that \ref{ta2}, \ref{ta3'} give us the expression for the trace of the static dark energy's energy-momentum tensor,
$$T^E=-\frac{(1+3w) \rho_{E,0}}{f^{\frac{1+w}{2w}}(r)}$$
Thus on the event horizon of a static black hole ($f(r)=0$), the trace diverges for $w<-1$ which implies the divergence of the Ricci scalar too, indicating a naked curvature singularity. Since the horizon of a static black hole corresponds to the infinities of the timelike Killing coordinate (e.g.~\cite{Wald:1984rg}), perhaps the above divergence is an analogue of the so called phantom disaster (e.g.,~\cite{Weinberg:2008zzc}) at cosmological late times for such dark energy models. }

.

\section{Time dependent cosmological scenario}
\subsection{$R_{\rm TA,max}$ in models with generic state parameter $w(t)$ }
\noindent We start with the  ansatz for the perturbative  McVittie spacetime, describing  the gravitational field of a point mass sitting in the ambient spatially homogeneous and isotropic cosmological spacetime with flat spatial sections, 
\begin{eqnarray}
ds^2=-(1+2\Psi(R,t))dt^2+ a^2(t)(1-2\Phi(R,t))\left(dx^2+dy^2+dz^2\right)
 \label{mv1}
\end{eqnarray}
where $\Psi$ and $\Phi$ are the weak and linearized gravitational potentials and $R=\sqrt{x^2+y^2+z^2}$ is the comoving radius. The above ansatz holds only for length scales much greater than the Schwarzschild radius of the central mass. In other words, it is suitable for describing large scale non-black hole structures such as the galaxies or  clusters of galaxies, whose Schwarzschild radii are located much inside them, implying  $|\Psi|,\, |\Phi| \ll1 $. A rather familiar example of such spacetimes would be the Schwarzschild-de Sitter spacetime written much outside the Schwarzschild radius, with $a(t)=e^{\sqrt{\Lambda/3} \;t}$ and $\Psi(R,t)=-2GM/Ra(t)=\Phi(R,t)$.

The Christoffel symbols for the above metric are
\begin{eqnarray}
\Gamma^{t}_{tt}&=&\dot{\Psi}, \quad \Gamma^t_{it}=\partial_i\Psi, \quad \Gamma^t_{ij}=\left[a\dot{a}(1-2\Phi-2\Psi)-a^2\dot{\Phi} \right]\delta_{ij}\nonumber\\
\Gamma^i_{tt}&=& \frac{\partial_i\Psi}{a^2}, \quad \Gamma^i_{jt}=(H-\dot{\Phi})\delta_{ij}, \quad \Gamma^i_{jk}=\partial_i\Phi \delta_{jk} -\partial_j\Phi \delta_{ik}-\partial_k\Phi \delta_{ij}
 \label{mv2}
\end{eqnarray}
where a `dot' denotes differentiation once with respect to time, $H=\dot{a}/a$ is the Hubble rate, the indices $(i,j)$ run between the spatial coordinates $x$, $y$ and $z$ and we have retained terms only linear in the potentials. The nonvanishing components of the  Ricci tensor are 
\begin{eqnarray}
R_{tt}&=&-3H^2 -3\dot{H}+\frac{\partial^2 \Psi}{a^2} +3\ddot{\Phi}+3H(\dot{\Psi}+2\dot{\Phi}), \quad R_{ti} =2\partial_i \left(\dot{\Psi}+H\Phi\right) \nonumber\\
R_{ij}&=&\left[(1-2\Phi-2\Psi)(a\ddot{a}+2\dot{a}^2)+\partial^2\Phi-a^2\ddot{\Phi}-a\dot{a}(\dot{\Psi}+6\dot{\Phi})\right]\delta_{ij}+\partial_i\partial_j (\Phi-\Psi) 
 \label{mv3}
\end{eqnarray}
where $\partial^2\equiv \partial_x^2+\partial_y^2+\partial_z^2$, whereas the components of the Einstein tensor are, 
\begin{eqnarray}
G_{tt}&=&3H^2 +\frac{2}{a^2}\partial^2\Phi-6H\dot{\Phi}, \quad G_{ti}=R_{ti} =2\partial_i \left(\dot{\Psi}+H\Phi\right) \nonumber\\
G_{ij}&=&\left[-\left(\frac{2\ddot{a}}{a}+H^2\right)+2(\Phi+\Psi) \left(\frac{2\ddot{a}}{a}+H^2\right)+\frac{\partial^2(\Psi-\Phi)}{a^2}+2 \ddot{\Phi}+2H(\dot{\Psi}+3\dot{\Phi})\right]a^2\delta_{ij} \nonumber \\
&&+\partial_i\partial_j (\Phi-\Psi)
\label{mv4}
\end{eqnarray}
We will use the above equations to do study cosmological scalar perturbation theory at the linear order with the dark energy and the cold dark matter as the sources. We shall ignore the backreaction effects due to any other matter fields such as the  electromagnetic radiation. The energy momentum tensor for the dark energy is given by, 
\begin{eqnarray}
T^E_{ab}=T^{E,0}_{ab}(t)+\delta T^{E}_{ab}(R,t)
 \label{mv5}
\end{eqnarray}
where the first term on the right hand side corresponds to the homogeneous background Friedmann-Robertson-Walker (FRW) spacetime,
\begin{eqnarray}
T^{E,0}_{ab}(t)= \rho_E(t)(\nabla_a t)(\nabla_b t) + P_E(t)a^2(t)\delta_{ab}  
 \label{mv6}
\end{eqnarray}
where the indices of the Kronecker delta run over the spatial coordinates only, while the energy density and the pressure satisfy
$$w(t)=P_E(t)/\rho_E(t)$$ 
The second term on the right hand side of \ref{mv5} corresponds to the inhomogeneous linear perturbation due to a central and spherical massive object, like a galaxy. Up to linear order the most general such perturbation should be written as
\begin{eqnarray}
\delta T^{E}_{ab}(R,t) \equiv  \left\{\delta \rho_E(R,t),~a^2 (t)\delta P_E (R,t) \delta_{ij},~a^2(t)\delta T^E_{ij}(R,t),~\delta T^E_{ti}(R,t)   \right\}  
 \label{mv7}
\end{eqnarray}
where we have ignored second order terms, e.g. of order $\sim \Psi \delta \rho_E$. $\delta T^E_{ij}$ is not proportional to $\delta_{ij}$ and, in particular, it may contain off-diagonal terms. Likewise,
we have the energy-momentum tensor corresponding to the cold dark matter 
\begin{eqnarray}
T^M_{ab}=\left[\rho(t) + M \delta^3(\vec{R}a(t))\right] (\nabla_a t)(\nabla_b t) +\delta T^M_{ab} (R, t)
 \label{mv8}
\end{eqnarray}
where the first term on the right hand side corresponds to the background FRW spacetime and the second term represents the central mass $M$ at rest with respect to the comoving frame. Since we are interested in determining the maximum size of a structure, the region of our interest is essentially outside it. Then, just outside the structure by virtue of the spherical symmetry of the problem, we may take all mass to be localized at $\vec{R}=0$. Finally, the third term in \ref{mv8} corresponds to a backreactionless or test spherical dark matter fluid whose dynamics we wish to study, 
\begin{eqnarray}
\delta T^M_{ab} (R, t)\equiv \left\{\delta \rho (R,t),~-\rho(t)a^2(t)v^i,~ {\cal O}(v^i\,v^j) \right\}
 \label{mv9}
\end{eqnarray}
where $v^i=dx^i/dt$ is the peculiar velocity of the test fluid element. Since that fluid under consideration is dark matter, it must be nonrelativistic i.e., $|v^i|\ll 1$ and hence we have ignored ${\cal O}(v^2)$
terms in the above ansatz. The negative sign in front of the second term is a mere convention -- after all we may flip  it to the positive sign  by just changing the direction of motion of the fluid. We further assume that the weak energy condition ($T_{tt}>0$) is satisfied by all matter fields and their perturbations. This completes the specification of the sources and we shall now use them into the Einstein equations. We further refer our reader to~\cite{Eingorn:2012dg, Eingorn:2015hza} for some recent developments of the cosmological perturbation theory with a positive $\Lambda$, around a central point mass. 

However, before proceeding any further, we may simplify the Einstein tensors in \ref{mv4} as follows. We note that, as we have repeatedly emphasized, we are interested here in a length scale much larger than the Schwarzschild radius and much smaller than the Hubble radius. In such scales, we may make a quasistatic approximation, in which the spatial derivatives of the potentials $\Psi$ and $\Phi$ are much larger than their temporal derivatives. Also, if $k^i$ is a component of the spatial momenta associated with the Fourier transformed potentials, it is clear from the length scale we are interested in that $|k^i|\gg H^{-1}(t)$. Putting these all in together and using \ref{mv6}-\ref{mv8}, we obtain for the Einstein equations,
\begin{eqnarray}
3H^2=8\pi G \left(\rho(t)+\rho_E(t)\right)\quad {\rm and }\quad \frac{\ddot{a}}{a}= -\frac{4\pi G}{3}\left[ \rho(t)+ \rho_E(t)(1+3w(t))\right]
\label{mv10}
\end{eqnarray}
for the FRW background and 
\begin{eqnarray}
&&\frac{1}{a^2(t)}\partial^2\Phi=4\pi G \left (M \delta^3(\vec{R}a(t))+\delta \rho_E(R,t)\right)\nonumber\\
&&\frac{1}{a^2(t)}\left[\partial^2(\Psi-\Phi)\delta_{ij}-\partial_i \partial_j (\Psi-\Phi)\right]=8\pi G \left[\delta P_E (R,t)\delta_{ij} + \delta T_{ij}(R,t) \right]\nonumber\\
&&\partial_i \left(\dot{\Psi}+H\Phi\right) = 4 \pi G \delta T^E_{ti} 
\label{mv11}
\end{eqnarray}
for the  perturbation. It is possible to solve, at least formally if not analytically, the set of \ref{mv10}, \ref{mv11} as follows. First, for a given dark energy
model, we can solve \ref{mv10} for the scale factor $a(t)$. Coming next to the first of the above equations, integrating both sides over a constant time hypersurface, using the divergence theorem and the spherical symmetry of the problem, it is easy to obtain  
\begin{eqnarray}
\frac{d\Phi}{dR}= \frac{G(M+\delta M_E(R,t))}{R^2 a(t)}
\label{mv12}
\end{eqnarray}
where $\delta M_E (R,t)$ is the contribution to the proper mass function due to the dark energy perturbation,
\begin{eqnarray}
\delta M_E(R,t):= \int a^3(t) \,R^2 \,dR \,d\Omega \, \delta \rho_E(R,t)
\label{mv13}
\end{eqnarray}
Since we have already assumed that all matter fields and their perturbation satisfies the weak energy condition, we have $\delta M_E (R,t) >0$. Note that the  mass function, $M+\delta M_E(R,t)$ could be thought of as an analogue of the Tolman-Oppenheimer-Volkoff mass function defined in a static scenario for a fluid~\cite{Wald:1984rg}, where the self gravitating fluid may now be imagined to have two components, both satisfying the positivity of the energy density.

Let us now come to the second of \ref{mv11}, in order to find the relation between the two potentials. Clearly $\Phi \neq \Psi$ only when $T^E_{ij}\neq 0$ for $i\neq j$~\cite{Mukhanov:2005sc}. Also, by the spherical symmetry of the problem, we have $\partial_{\theta}(\Phi-\Psi)=0=\partial_{\phi} (\Phi-\Psi)$. Then, converting the indices $(i,j)$ from Cartesian to spherical polar, we have $\delta T^E_{\theta \theta}=\delta T^E_{\phi\phi}=\delta T^E_{\theta \phi}=0$, and
\begin{eqnarray}
 \frac{d^2(\Phi-\Psi)}{dR^2}= 8\pi \,G \,a^2(t)\, \delta T^E_{RR}(R,t)
\label{mv14}
\end{eqnarray}
For the explicit form of the above equation in the case of Brans-Dicke cosmology we refer our reader to~\cite{Bhattacharya:2016vur}. Using \ref{mv13}, we now obtain 
\begin{eqnarray}
 \frac{d \Psi}{dR}=\frac{G (M +\delta M_E)}{R^2 a(t)} -8\pi G a^2(t) \int dR\, \delta T^E_{RR}(R,t) \equiv \frac{G M_{\rm tot}(R,t)}{R^2 a(t)}
\label{mv15}
\end{eqnarray}
where we have set the integration constant to zero, as in the absence of any inhomogeneity, we must have $\Psi=1$.  We also have  defined a total or effective mass function,
\begin{eqnarray}
 M_{\rm tot}:= M+\delta M_E-8\pi  a^3(t)R^2 \int dR\, \delta T^E_{RR}(R,t)
\label{mv16}
\end{eqnarray}
Let us now come to the sector of the cold dark matter. The homogeneous part, $\rho(t)$ satisfies the conservation equation
$$\dot{\rho}(t)+3H \rho(t)=0$$
whereas the inhomogeneous backreactionless perturbation specified in \ref{mv9} satisfies (the conservation equation $\nabla^a\delta T^M_{ab}=0$, in the background~\ref{mv1},
\begin{eqnarray}
&&\delta{\dot \rho}+\rho \partial_i v^i+3 H \delta \rho=0\nonumber\\
&&\dot{v}^i+2H v^i +\frac{\partial_i \Psi}{a^2}=0.
\label{mv17}
\end{eqnarray}
The maximum turn around radius is defined as the point where the proper `peculiar acceleration' $(=d^2 (a(t)x^i)/dt^2)$ of a fluid element vanishes which, after using the second of \ref{mv17} becomes
\begin{eqnarray}
\frac{\ddot{a}\,x^i}{a}-\frac{\partial_i \Psi}{a^2}=0.
\label{mv18}
\end{eqnarray}
As discussed in the previous section, the largest of all the maximum turn around radii is obtained for zero orbital angular momentum. Thus, setting $i=R$ above and using \ref{mv10},~\ref{mv15}, we obtain 
\begin{eqnarray}
-\frac{4\pi G \rho_E(t)}{3}  (1+3w(t))R-\frac{G M_{\rm tot}}{R^2 a^3(t)} -\frac{4\pi G \rho (t)}{3}R=0.
\label{mv19}
\end{eqnarray}
Writing $\delta M := \frac43 \pi \rho(t) R^3 a^3(t)$ and $\Lambda(t) :=\rho_E(t)/8\pi G$, we find the proper maximum turn around radius
\begin{eqnarray}
R_{\rm TA,max}= a(t)R\big\vert_{\rm max}=\left(-\frac{6 G M_{\rm eff}(t)}{\Lambda(t)(1+3w(t))} \right)^{\frac13},
\label{mv20}
\end{eqnarray}
where we have defined an effective mass function $M_{\rm eff }= M_{\rm tot}+\delta M$, taking into account the contribution of all sources. The above expression for the maximum turn around radius  is formally similar to the result of~\cite{PavlidouTetradisTomaras2014} obtained for a constant equation of state parameter and without considering any dark energy perturbation, whose effects here have been dumped into the mass function~$M_{\rm eff}$. Setting $w(t)=-1$ in~\ref{mv20} one recovers the $\Lambda{\rm CDM}$ result.  Note also that in deriving the above expression, no assumption, whatsoever was made on the behaviour of $w(t)$. Thus, from the analysis of~\cite{PavlidouTetradisTomaras2014}  it follows that {\it any} dark energy model falling into the generic class we are considering here with $w\,(t=t_{\rm today})\gtrsim -2.3$ is consistent with the structure size vs mass observational data.

A few of the  geometric aspects of the maximum turn around radius could be noted here. Firstly and obviously, the proper  radial distance which is purely spatial, is an invariant owing to the  spatial isotropy of the systems like \ref{ta1}, \ref{mv1},~\cite{Weinberg:2008zzc}. In other words, for all observers obeying the isometries, $R_{\rm TA, max}$ is an invariant quantity. Second, we may note here the purely geometric origin  of the definition of the turn around radius in Sec.~2, from the vanishing of the acceleration of the orbits of the timelike Killing vector field. For a discussion on how such consideration in a static spacetime \ref{ta1}, could be linked to the general McVittie spacetime~\ref{mv1}, we refer our reader to~\cite{Bhattacharya:2016vur}.

Before ending this section, we wish to show that \ref{mv18} could in fact be related to the geodesic equation. Indeed, let us first rewrite the metric \ref{mv1} in terms of the proper radial coordinate
\begin{eqnarray}
r:= (1-\Phi(R,t)) a(t) R
\label{mv21}
\end{eqnarray}
Since we are much outside the Schwarzschild radius of the structure and much inside the Hubble radius ($=H^{-1}(t)$), the metric becomes   to linear order in $\Psi,~\Phi$ and $H(t)r$
\begin{eqnarray}
ds^2= -\left(1+2\Psi-H^2(t)r^2\right)dt^2-2H(t)r dr dt+(1+2r\Phi')dr^2 +r^2 d\Omega^2
\label{mv22}
\end{eqnarray}
where the prime denotes differentiation once with respect to the proper radial coordinate. Let us consider a test particle moving along a timelike geodesic in the above geometry,
$$\frac{d^2 x^{\mu}}{d\lambda^2}+\Gamma^{\mu}_{\nu\rho}\frac{d x^{\nu}}{d\lambda}\frac{d x^{\rho}}{d\lambda} =0 $$
where $\lambda$ is an affine parameter and we consider a radial, nonrelativistic trajectory  as earlier. In this limit, $\lambda \approx t$ and hence $dt/d\lambda \sim 1 $. Thus the conservation of norm of the geodesic implies : $dr/d\lambda \ll 1$. Then we have, 
\begin{eqnarray}
\frac{d^2r}{dt^2}+\left(g^{rr}\partial_t g_{tr}+\frac12 g^{tr}\partial_t g_{tt}-\frac12 g^{rr}\partial_r g_{tt} \right) \approx 0
\label{mv23}
\end{eqnarray}
Using \ref{mv22} and its inverse and keeping in mind that the radial derivative of the potential dominates over the temporal one at the length scales we are interested in, we obtain
\begin{eqnarray}
\frac{d^2r}{dt^2}-\dot{H}(t)r-H^2(t)r+\Psi' =0
\label{mv24}
\end{eqnarray}
Inserting into this the vanishing acceleration condition, $d^2r/dt^2=0$, \ref{mv18} is reproduced. Clearly, such equivalence between the perfect fluid and geodesic is possible due to the fact that to derive \ref{mv18}, we had ignored all higher order terms coming from the energy-momentum tensor of the test fluid.

\subsection{Applications of the above formalism}
\bigskip

\subsubsection{Quintessence models}
\noindent
As an application of the above general formalism, 
the first example we take is the {\it quintessence model}, originally proposed in~\cite{Ratra} to address the coincidence problem of the dark energy (see also~\cite{Tsujikawa:2013fta} for a review). The role of the dark energy is played here by a time dependent scalar field $\phi_0(t)$. The model is described by the action
\begin{eqnarray}
S= \int d^4x \sqrt{-g} \left[\frac{R}{16\pi G} -\frac12 (\nabla_a\phi)(\nabla^a\phi)-V(\phi)+{\cal L}_M\right]
\label{ex1}
\end{eqnarray}
where $V(\phi)$ is some positive potential, to be chosen appropriately in order to generate the accelerated expansion
and ${\cal L}_M$ is the Lagrangian density corresponding to any other matter field, such as the cold dark matter.
Assuming a homogeneous FRW background solution with $\phi=\phi_0(t)$, its energy density and pressure are given by
\begin{eqnarray}
\rho_E(t)=\frac{\dot{\phi_0}^2}{2}+V(\phi_0) \quad {\rm and}  \quad P_E(t)=\frac{\dot{\phi_0}^2}{2} -V(\phi_0)
\label{ex2}
\end{eqnarray}
respectively.
 
There could be various choices of the potential $V(\phi)$. For example, the so called {\it tracker model},
\begin{eqnarray}
V(\phi)={\cal M}^{4+p}\phi^{-p}
\label{ex4}
\end{eqnarray}
where the parameter ${\cal M}$ has dimensions of mass and $p$ is a positive real number. Or, the so called {\it thawing model} :
\begin{eqnarray}
V(\phi)= {\cal M}^4\left(1+\cos(\phi/\mu) \right)
\label{ex5}
\end{eqnarray}
with the parameters ${\cal M}$ and $\mu$ having dimensions of mass. For a review on different such choices for the potential and their cosmological implications including the perspective of particle physics, we refer our reader to~\cite{Tsujikawa:2013fta} and references therein. 

For any potential $V(\phi)$ the homogeneous background is a fluid of the type studied in the previous section, with state `parameter' given by
\begin{eqnarray}
w(t)= \frac{\dot{\phi_0}^2-2V(\phi_0)}  {\dot{\phi_0}^2+2V(\phi_0)}
\label{ex3}
\end{eqnarray}
Thus, it is clear that in order  to have $w(t)<0$, the potential must dominate over the kinetic term. Furthermore, for positive potential the state parameter satisfies $w(t)\geq -1$. Then, according to the discussion of this section, one may immediately conclude that the generic model of this class predicts cosmic structure sizes greater than $\Lambda{\rm CDM}$ and, thus, is consistent with the structure size vs mass observations.

\bigskip

\subsubsection{The generalized Chaplygin gas model}

\noindent
The other theory we wish to address here is the {\it generalized Chaplygin gas model}. The Chaplygin gas  or its generalized version attempts to give a unified picture of the dark energy and the dark matter~\cite{Bento:2002ps, Bento:2003dj}. Let us consider the equation of state
\begin{eqnarray}
  P= -\frac{A}{\rho^{\beta}},
 \label{de5}
\end{eqnarray}
where both $\beta$ and $A$ are positive constants. The conservation equation : $\dot{\rho}+3H (\rho+P)=0$  gives
\begin{eqnarray}
\rho= \left[A+ \frac{\rho_m^{\beta+1}}{a^{3(\beta+1)}}\right]^{\frac{1}{\beta+1}}
 \label{de6}
\end{eqnarray}
where $\rho_m$ is a positive integration constant. At sufficiently late times, $a(t) \gg1$ (the current scale factor is set to unity), we have $\rho \approx A^{\frac{1}{\beta+1}}$ and hence $P\approx - A^{\frac{1}{\beta+1}}$, giving the equation of state of a cosmological constant. On the other hand at sufficiently early times, $a(t) \ll 1$ and we have $\rho \sim a^{-3}(t)$ and $|P| \ll \rho$, i.e. the matter dominated universe. 

The original Chaplygin gas model was defined specifically for $\beta=1$. Then, it turns out that the dust dominated universe is smoothly interpolated with the dark energy dominated universe via an intermediate equation of state, $P=\rho$, known as the {\it stiff matter}. Such an intermediate equation of state is necessary in order to ensure that the normal matter has positive pressure.
The generalized Chaplygin gas model, on the other hand,  permits $0<\beta \leq 1$, which we shall consider below. In this case the matter and dark energy dominated universe is interpolated via an equation of state $P=\beta \rho$ with $\beta>0$ being a parameter of the theory and $\beta=0$ corresponding to the $\Lambda{\rm CDM}$.

We are interested in the current universe dominated by the dark energy. So, we may expand \ref{de5} and \ref{de6} to obtain at leading order
\begin{eqnarray}
\rho \approx  A^{\frac{1}{1+\beta}}+\frac{ \rho_0^{1+\beta} }{(1+\beta) A^{ \frac{\beta}{1+\beta} } a^{3(1+\beta)}}, \qquad P \approx  -A^{\frac{1}{1+\beta}}+\frac{\beta \rho_0^{1+\beta} }{(1+\beta) A^{\frac{\beta}{1+\beta} } a^{3(1+\beta)}}
 \label{de7}
\end{eqnarray}
Thus, the above two equations represent the dynamics of our universe with an effective homogeneous fluid with equation of state $P(t)=-A^{\frac{1}{1+\beta}}+\beta (\rho (t)-A^{\frac{1}{1+\beta}})$. Solving Eqs.~(\ref{mv10}) for the scale factor we obtain
\begin{eqnarray}
a(t) \sim \left(\sinh \frac{3{\cal H}_0 (1+\beta ) t}{2}\right)^{\frac{2}{3(1+\beta )}},
 \label{de8}
\end{eqnarray}
with ${\cal H}_0 = \sqrt{8\pi G A^{1/(1+\beta)}/3 }$. For $t\to \infty$, we recover the de Sitter space, while for $\beta=0$ we recover the $\Lambda{\rm CDM}$. In order for this model to be consistent with  modern cosmology, the value of the parameter $\beta$ needs to be rather small, $\beta \lesssim 0.05$~\cite{Marttens:2017njo}.

\ref{de7} show that we do not need to consider the inhomogeneous dark energy perturbation in this case, and hence we have $\Psi=\Phi$ in \ref{mv1}. With this, we study the dynamics of a
test  Generalized Chaplygin gas fluid, following the method described in the preceding section. Since the parameter $\beta$ is small, we take terms like $\beta \delta \rho$ or $\beta v^i$ to be second order only, giving the maximum turn around condition, 
\begin{eqnarray}
\frac{\ddot{a}}{a} R -\frac{1+\beta}{a^2}\partial_R \Psi =0
 \label{de15}
\end{eqnarray}
The Chaplygin gas equation of motion gives,
$$\frac{\ddot{a}}{a}=\frac{8\pi G A^{1/(1+\beta)}}{3}-\frac{4\pi G (1+3\beta) \rho(t)}{3} $$
Writing $8\pi G A^{1/(1+\beta)}= \Lambda$ and following the procedure described in the previous section, it is straightforward to  obtain to first order in the parameter $\beta$, 
\begin{eqnarray}
R_{\rm TA,max}=\left(\frac{3 G M_{\rm eff}}{\Lambda} \right)^{\frac13}\left(1+\frac{\beta}{3} +\frac{2\beta \delta M}{M_{\rm eff}} \right)
 \label{de16}
\end{eqnarray}
where $\delta M := \frac43 \pi R^3 a^3(t) \rho(t) $ as earlier. Setting $\beta=0$ one recovers the $\Lambda{\rm CDM}$ result. Thus, since $\beta>0$ the prediction of the generalized Chaplygin gas model for the maximum turnaround radius is larger than in $\Lambda{\rm CDM}$. This fact can be understood as the effect of the positive pressure fluid, that opposes the accelerated expansion and implies that the generalized Chaplygin gaz too is consistent with the data on the sizes and the masses of the cosmic structures.   

\section{Conclusions}
In this work we have discussed various aspects of the maximum possible structure size, in the context of the mass versus observed sizes of some nearby large scale cosmic structures. In Sec.~2, we have given a  derivation of the  maximum turn around radius for a dark energy with constant equation of state parameter ($P=w\rho$ with $w$ const.) in a purely static geometry derived at the maximum turn around length scale (\ref{ta7'}). This result was found earlier in~\cite{PavlidouTetradisTomaras2014} using cosmological perturbation theory. Even though for 
$w\neq -1$, the dark energy is usually taken to be time dependent and spatially homogeneous and isotropic, we have shown here that the same expression for the $R_{\rm TA,max}$ holds also when we consider the dark energy to be inhomogeneous and time independent.  

In Sec.~3 we have derived the expression for the $R_{\rm TA,max}$ in a cosmological scenario, using the McVittie geometry in the weak field limit. The dark energy we have taken here is generic and has completely {\it arbitrary, time dependent } state `parameter' ($P(t)=w(t)\rho(t)$ with $w(t)$ arbitrary). We also have taken the  first order backreaction effect on the dark energy due to the central inhomogeneity. The final result obtained in \ref{mv20} is formally exactly similar to that of the case of $w(t)={\rm const.}$ and hence giving the same bound on the numerical value of the state parameter, $w\,(t=t_{\rm today})>-2.3$. We have applied this general formalism in Sec.~3.2 to two quite popular alternative dark energy/gravity models viz. the quintessence and the generalized Chaplygin gas and have shown that both these models pass the size versus mass test with flying colours.

The framework  we have considered, even though it goes beyond the $\Lambda {\rm CDM}$, essentially is described by the Einstein equations. It should thus be highly interesting to extend such general framework  to go beyond the usual 4-dimensional Einstein equations and to address more complicated   scenarios such as the brane-world~\cite{Bag:2016tvc}. We hope to return to this issue in the near future.

\section*{Acknowledgements}
\noindent
We would like to thank V.~Pavlidou and C.~Skordis for useful discussions.


\end{document}